\documentclass{webofc}
\usepackage[varg]{txfonts}   
\usepackage{hyperref}
\usepackage{xcolor}
\usepackage{amsmath}
\begin{document}
\title{An updated and improved thermal SZ $y$-map from \textit{Planck} PR4 data }
%
%

\author{\lastname{Jyothis Chandran}\inst{1}\fnsep\thanks{\email{chandran@ifca.unican.es}} \and
        \lastname{Mathieu Remazeilles}\inst{1} \and
        \lastname{R. B. Barreiro}\inst{1} 
}

\institute{Instituto de Física de Cantabria (CSIC-UC), Avda. de los Castros s/n, 39005 Santander, Spain}

\abstract{%
  In 2015, the \textit{Planck} Collaboration released an all-sky map of the thermal Sunyaev-Zeldovich (SZ) effect, obtained by implementing the needlet internal linear combination (NILC) method on the \textit{Planck} PR2 data. The quality of the \textit{Planck} data has significantly improved since then. The \textit{Planck} PR4 data release offers upgraded full-sky maps in the LFI and HFI frequency bands with improved systematics and sensitivity. We present a new all-sky thermal SZ Compton $y$-parameter map derived from the \textit{Planck} PR4 data using NILC and highlight improvements, particularly in noise reduction and handling residual foreground contamination. The PR4 NILC Compton $y$-parameter map has been made publicly available to support further analyses.

}
\maketitle
%
\section{Introduction}
\label{sec:intro}

Cosmic microwave background (CMB) photons, on their journey from the last-scattering surface to us, experience inverse Compton scattering when they interact with high-energy electrons within hot ionized gas regions along the way. This process occurs not only within galaxy clusters but also in the unbound gas regions between clusters. The CMB photons, on average, gain energy from these high-energy electrons, resulting in a shift toward higher frequencies in their energy distribution. This is visible as a spectral distortion of the CMB blackbody spectrum and is called the thermal Sunyaev-Zeldovich (SZ) effect \cite{Zeldovich:1969}. 
It is possible to produce a Compton $y$-parameter map (or $y$-map), tracing the amplitude of the thermal SZ effect across the sky, by using weighted linear combinations of multifrequency sky observations.
Such a map can serve as a direct tracer of the hot baryons within and between galaxy clusters and an indirect tracer of the dark matter distribution. Either alone or in combination with primary CMB or large-scale structure observations, the thermal SZ $y$-map can provide valuable constraints on cosmological parameters such as the amplitude of dark matter fluctuations, $\sigma_8$, and the matter density, $\Omega_{\rm m}$.

\begin{figure}[h]
\centering
\sidecaption
\includegraphics[scale=0.15]{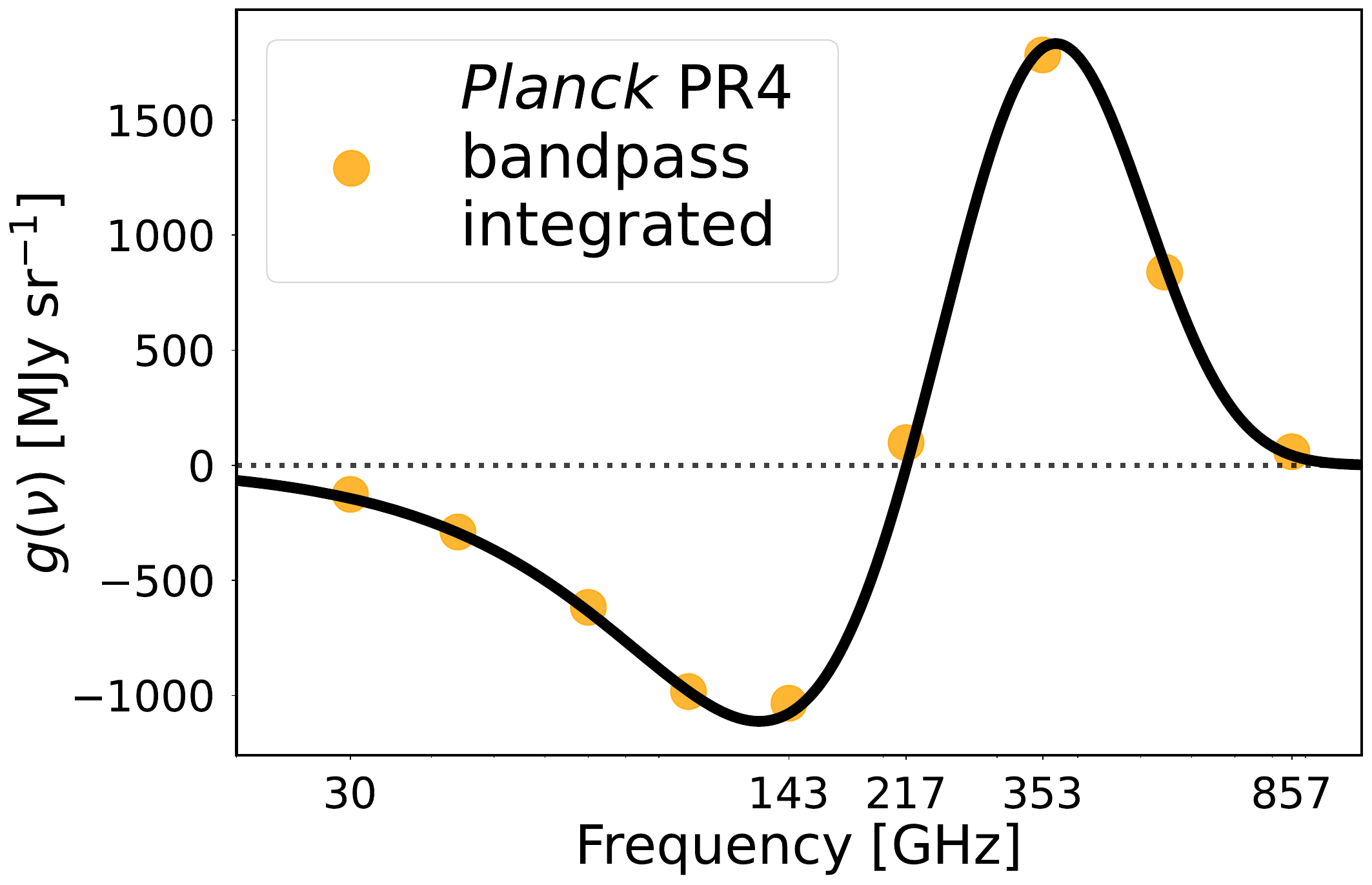}
\caption{The frequency response, or spectral energy distribution (SED), of the thermal SZ effect in intensity units and the PR4 bandpass integrated coefficients corresponding to the \textit{Planck} frequency channels. The frequency channels of the \textit{Planck} instruments trace the frequency response of the thermal SZ across its negative peak ($\sim$ 140 GHz), null ($\sim$ 220 GHz), and positive peak ($\sim$ 350GHz).}
\label{fig:freq}       
\end{figure}

The \textit{Planck} satellite mission provided full-sky coverage across multiple frequency channels. Notably, the nine \textit{Planck} frequencies ranging from 30 to 857 GHz are distributed to effectively capture the spectral response of the thermal SZ effect (see Fig.~\ref{fig:freq}). Therefore, using the \textit{Planck} data release 2 (PR2, \cite{PR2:Plk2015I}), two $y$-maps were released in 2015 by \cite{Remazeilles:Plk2015XXII}, generated using two different implementations, NILC and MILCA, of the internal linear combination (ILC) method to combine the frequency maps with optimal weighting. Subsequently, the quality of the \textit{Planck} data has improved, culminating in the \textit{Planck} data release 4 (PR4) \cite{NPIPE:Plk2020LVII}. In this proceedings paper, we present a new improved $y$-map derived from \textit{Planck} PR4 data using the NILC (needlet internal linear combination, \cite{NILC2009}) method, as detailed in \cite{NILC_PR4_ymap}. 

\section{Reconstruction of the Compton $y$-map} 
\label{sec:methodology}

The \textit{Planck} observations consist of full-sky maps across nine frequency bands, each of them containing a combination of the thermal SZ signal, the CMB, various astrophysical foreground emissions, and instrumental noise. Extracting and mapping the faint thermal SZ signal with high fidelity is a real component separation challenge.

\subsection{\textit{Planck} Release 4 (PR4) data}
\label{subsec:pr4}

In this work, we use the nine full-sky frequency maps from the \textit{Planck} PR4 release \cite{NPIPE:Plk2020LVII}. This data set has lower instrumental noise levels compared to earlier releases, primarily due to $\sim$8\% additional data from repointing manoeuvre. Enhanced destriping using a short baseline has effectively reduced the striping artefacts observed in the scanning direction of the \textit{Planck} satellite in earlier data releases. Unlike previous releases, the Low-Frequency Instrument (LFI) and High-Frequency Instrument (HFI) PR4 data were processed coherently within a single pipeline (NPIPE). Calibration and HFI frequency bandpasses also exhibit differences in this release. It is worth mentioning that the Solar dipole and a frequency-dependent kinematic quadrupole component were retained in the PR4 frequency maps, requiring subtraction before component separation. In addition to the full-mission PR4 maps, we use the PR4 half-ring (HR) data splits to generate two supplementary $y$-maps (HR1 and HR2 $y$-maps) for noise characterization.

\subsection{NILC implementation}
\label{subsec:nilc}

In this study, we implement the NILC method \cite{NILC2009}, following the procedure described in \cite{Remazeilles:Plk2015XXII}, with some differences as outlined below. As an ILC method, NILC takes advantage of the known spectral response of the tSZ signal to perform a constrained weighted linear combination of the frequency maps that preserves the signal while minimizing the variance of the foregrounds. Furthermore, being a \emph{wavelet-based} ILC, NILC enables localized filtering in both pixel and harmonic domains simultaneously, with the NILC weights adjusting dynamically based on changing foreground contamination conditions across the sky and different angular scales.

We perform NILC across 10 multipole ranges, using Gaussian-shaped \emph{needlet} (spherical wavelet) windows. A small mask covering $\sim$2\% of the PR4 maps is applied before NILC processing to prevent ringing effects due to the brightest pixels in the Galactic centre. The non-Gaussian beams of the PR4 data are used to deconvolve the effect of the instrument beam across frequencies, a departure from the PR2 $y$-map reconstruction, which employed Gaussian beam approximations. The choice of \textit{Planck} frequency channels in NILC is selective and depends on the specific range of angular scales. This strategy prevents the low-resolution LFI channels from contributing to the small scales in the $y$-map and mitigates the contamination from the cosmic infrared background (CIB) by excluding the $857$\,GHz channel at those scales.
For more on methodology, see section 3.2 of \cite{NILC_PR4_ymap}.

\section{Results} \label{sec:results}

\subsection{Map inspection} \label{subsec:maps}

\begin{figure}
\centering
\includegraphics[scale=0.3]{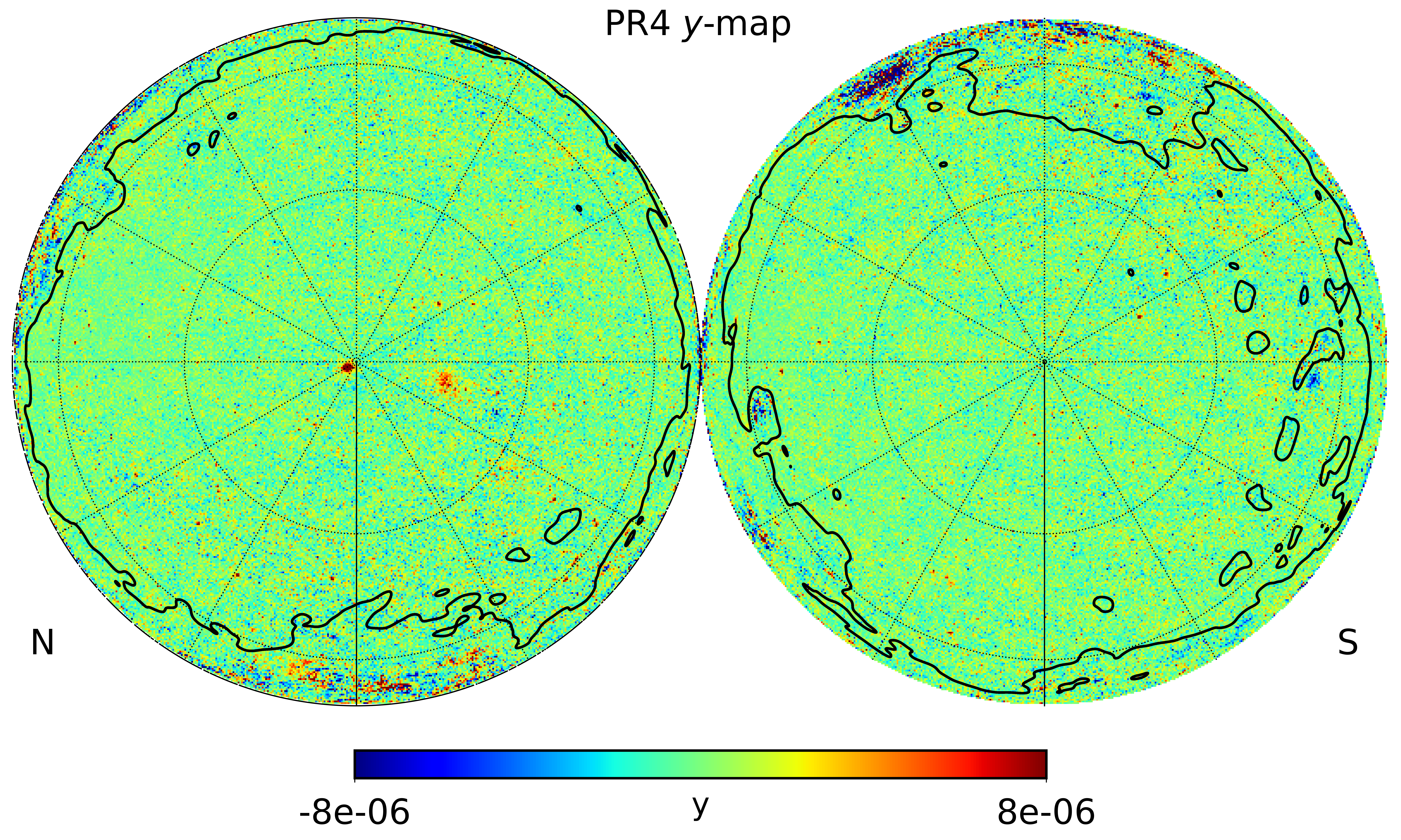}
\caption{The NILC $y$-map from \textit{Planck} PR4 data in orthographic projection (10' resolution). The left side corresponds to the northern hemisphere, and the right side corresponds to the southern hemisphere in Galactic coordinates. The black line outlines the boundary of the Galactic mask used to mitigate residual Galactic foreground contamination when computing power spectra and other statistics.}
\label{fig:ymap}       
\end{figure}

Figure~\ref{fig:ymap} shows the new PR4 $y$-map produced with NILC. The thermal SZ signal from galaxy clusters is visible as red spots, with examples like Coma and Virgo visible in the central region of the northern hemisphere. Along the edges, residual Galactic foreground contamination is noticeable. In comparison to the PR2 $y$-map, the new PR4 $y$-map exhibits lower granularity due to reduced noise and extragalactic foreground contamination. Further details about this improvement are discussed in the subsequent sections. Moreover, the PR4 $y$-map lacks the patchy blue artefacts near the bottom of the southern hemisphere observed in the PR2 version.

\begin{figure}
\centering
\includegraphics[scale=0.233]{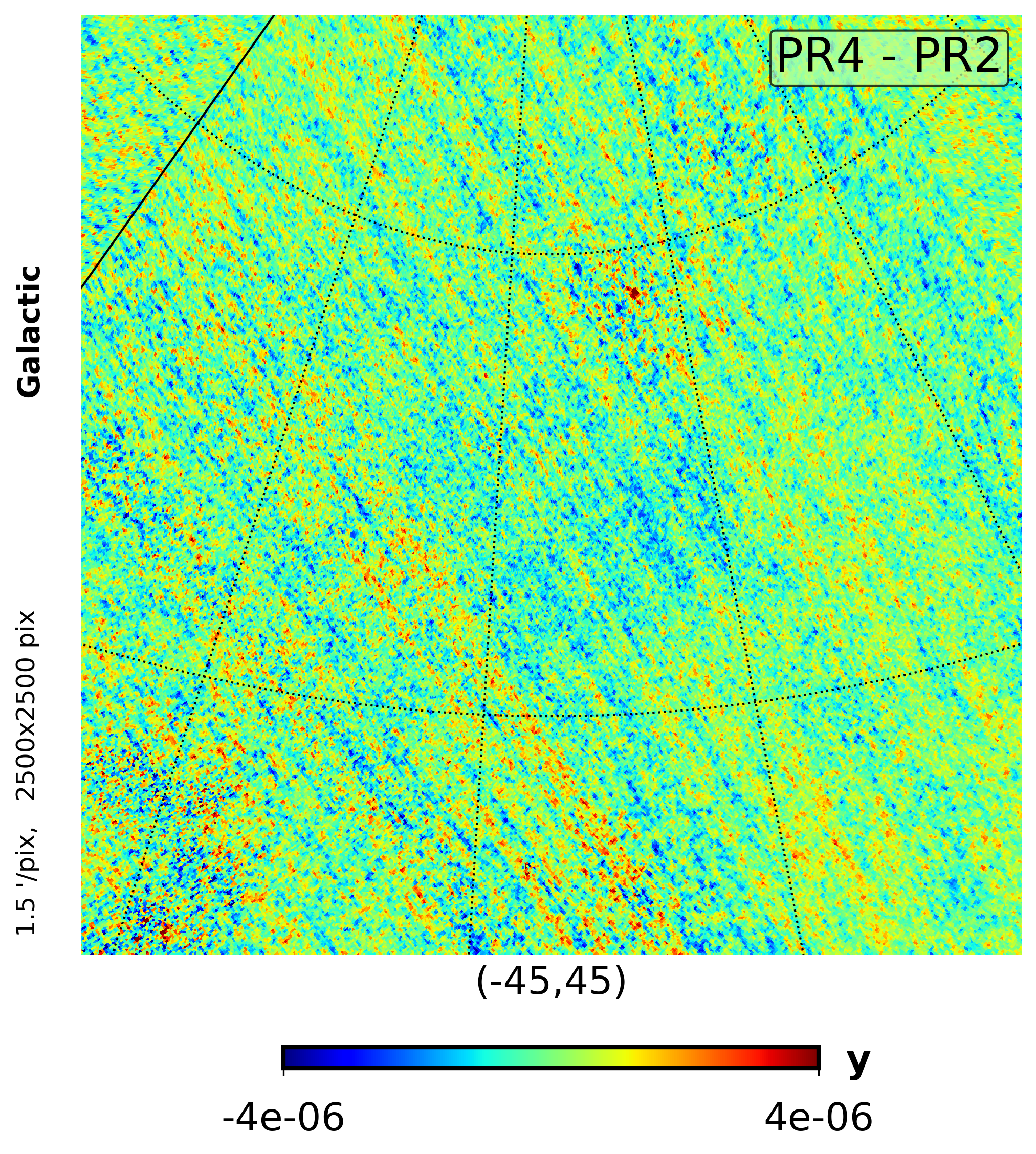}
\hfill
\includegraphics[scale=0.205]{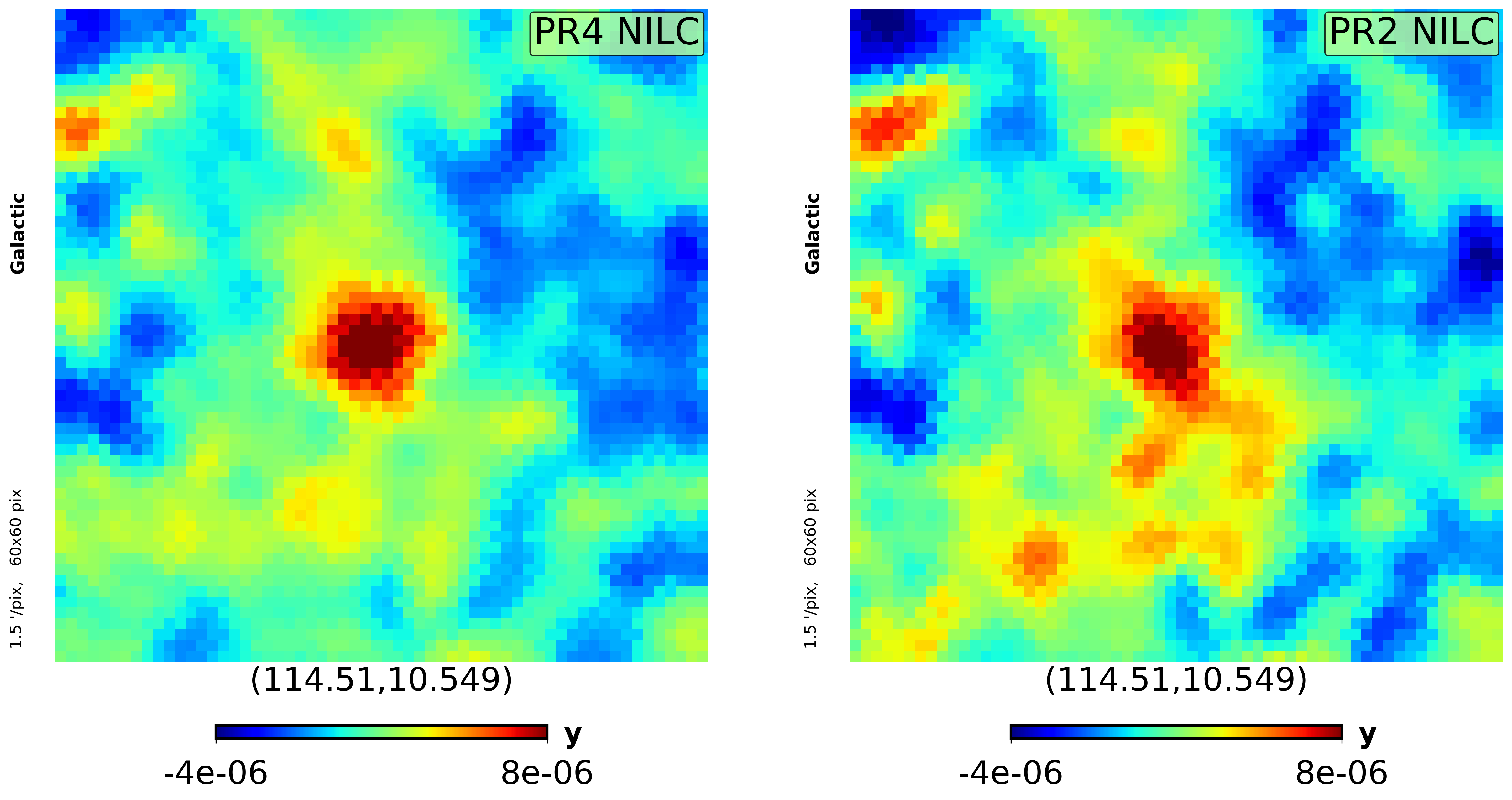}
\caption{The difference between the PR4 and PR2 $y$-maps (\emph{left}) in a large sky area ($\sim60^\circ\times 60^\circ$). Residual stripes from the PR2 $y$-map along the \textit{Planck} scanning direction are observable. Low Galactic latitude galaxy cluster CIZA J2302.7+7137 in the PR4 (\emph{middle}) and PR2 (\emph{right}) NILC $y$-maps.}
\label{fig:ymap_regions}       
\end{figure}

The PR2 $y$-maps were known to have  significant residual large-scale striations along the scanning direction due to $1/f$ noise (see section 4.1 of \cite{Remazeilles:Plk2015XXII}). Even though the striations are large-scale features, they can change the brightness of compact SZ sources along their direction. Thanks to an improved destriping strategy implemented in the PR4 frequency maps by NPIPE, the resulting PR4 $y$-map exhibits reduced levels of striping in comparison to the PR2 $y$-maps. This is illustrated in Figure~\ref{fig:ymap_regions}, where we show the difference between the PR4 and PR2 NILC $y$-maps (\emph{left panel}), highlighting the contrast in striping levels between the two $y$-maps. 
Figure~\ref{fig:ymap_regions} also highlights a galaxy cluster (CIZA J2302.7+7137 \cite{CIZA}) positioned at low Galactic latitude near the Galactic plane in both the PR4 (\emph{middle panel}) and PR2 (\emph{right panel}) NILC $y$-maps. The cluster exhibits improved resolution and morphology in the new map.

\begin{figure}
\centering
\includegraphics[scale=0.233]{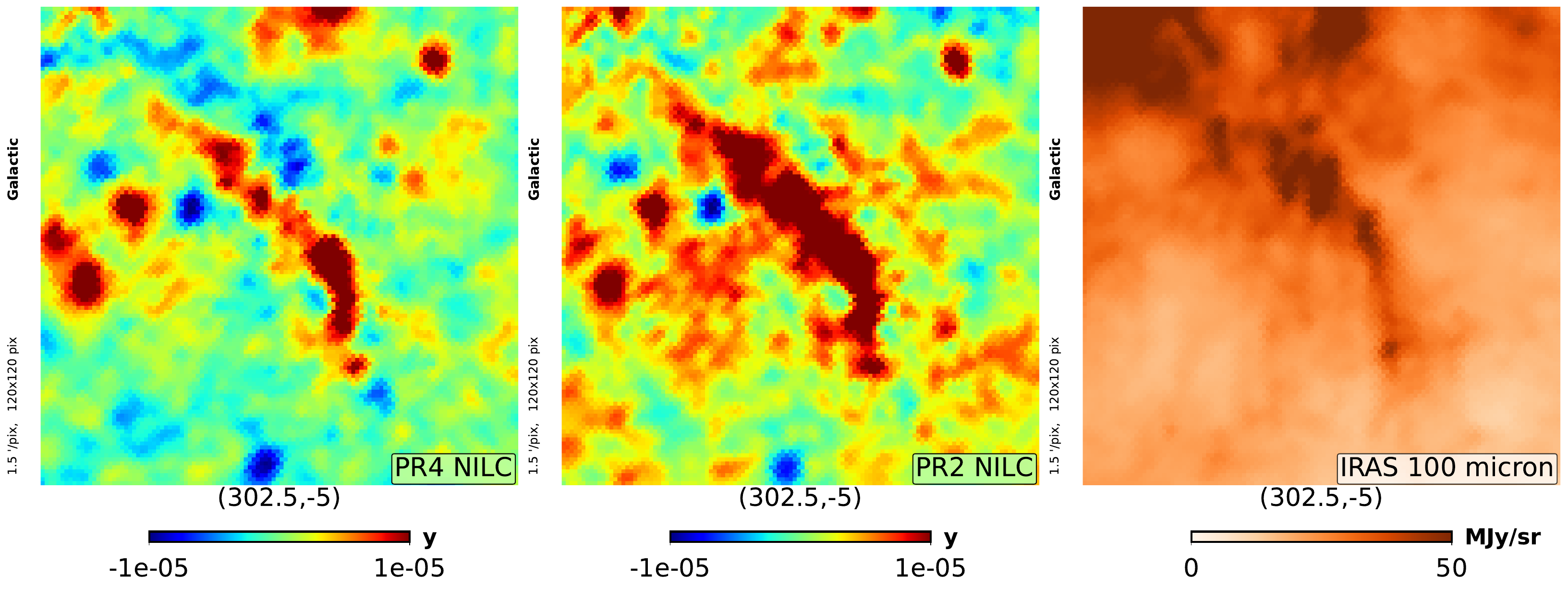}
\caption{A $3^\circ\times 3^\circ$ region near the Galactic plane featuring a dust filament in the PR4 $y$-map (\emph{left}), the PR2 $y$-map (\emph{middle}), and the IRIS (reprocessed IRAS) 100-micron map (\emph{right}). The PR4 $y$-map exhibits significantly reduced residual Galactic dust contamination in comparison to the PR2 $y$-map.}
\label{fig:ymap_dust}       
\end{figure}

Galactic thermal dust emission acts as a major foreground contaminant at large angular scales and low Galactic latitudes. As evident from Figure \ref{fig:ymap_dust}, the PR2 $y$-map (\emph{middle panel}) exhibits significant residual contamination from a thermal dust filament identified in the IRIS 100-micron map (\emph{right panel}), which is notably reduced in the PR4 $y$-map (\emph{left panel}).
The RMS (root mean square) of the PR4 $y$-map within the Galactic region outlined by the black line in Figure~\ref{fig:ymap}  is also 5\% lower than that of the PR2 $y$-map. This corroborates the fact that the new map has lower contamination from Galactic foregrounds.

\subsection{One-point statistics (1-PDF)}
\label{subsec:pdf}

\begin{figure}
\sidecaption
\centering
\includegraphics[scale=0.115]{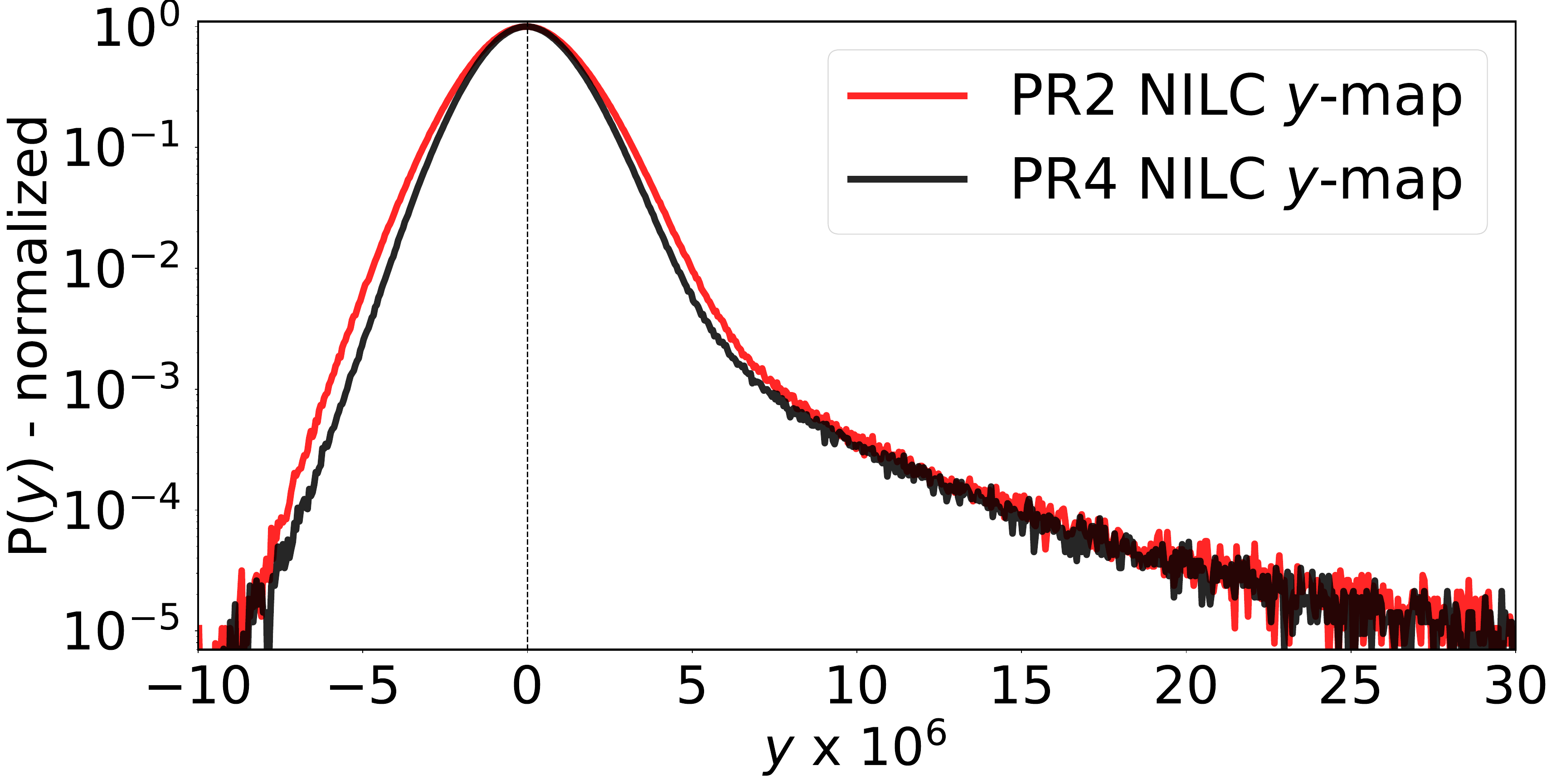}
\caption{The 1-PDF of the PR2 (red) and PR4  (black) NILC $y$-maps. The positive tail characteristic of thermal SZ emission is consistent between the two $y$-maps, while the width (variance) of the distribution originating from contaminants is reduced for the PR4 $y$-map.}
\label{fig:1pdf}       
\end{figure}

The thermal SZ effect has a characteristic non-Gaussian distribution, with a positively skewed tail \cite{sz_skewness}. Figure~\ref{fig:1pdf} shows the one-point probability density function (1-PDF) of the PR2 $y$-map (red) and the PR4 $y$-map (black), obtained by computing the histogram of the maps over $\sim 56\%$ of the sky after masking the Galactic region (outlined in black in Figure~\ref{fig:ymap}) and the point-sources detected at frequencies below $217$\,GHz. The point-source mask has been constructed at each frequency channel using the Mexican Hat Wavelet 2 \cite{point_sources, marcos} as part of the Sevem pipeline \cite{NPIPE:Plk2020LVII}. Both $y$-maps display consistent positive tails, indicating a coherent reconstruction of the signal.

However, the width of the two distributions differs due to lower noise and foreground contamination in the PR4 $y$-map, resulting in a $17\%$ lower variance in the 1-PDF compared to the PR2 $y$-map. To estimate the noise contribution to the 1-PDF of the PR2 and PR4 $y$-maps, we also computed the 1-PDF of the half-difference of $y$-maps reconstructed from half-ring data splits (HR1 and HR2 $y$-maps, as described in section \ref{subsec:pr4}). The analysis revealed a $7\%$ reduction in the variance of the noise 1-PDF for the PR4 $y$-map compared to the PR2 $y$-map.


\subsection{Two-point statistics (angular power spectrum)}
\label{subsec:cl}

\begin{figure}[h]
\centering
\includegraphics[scale=0.11]{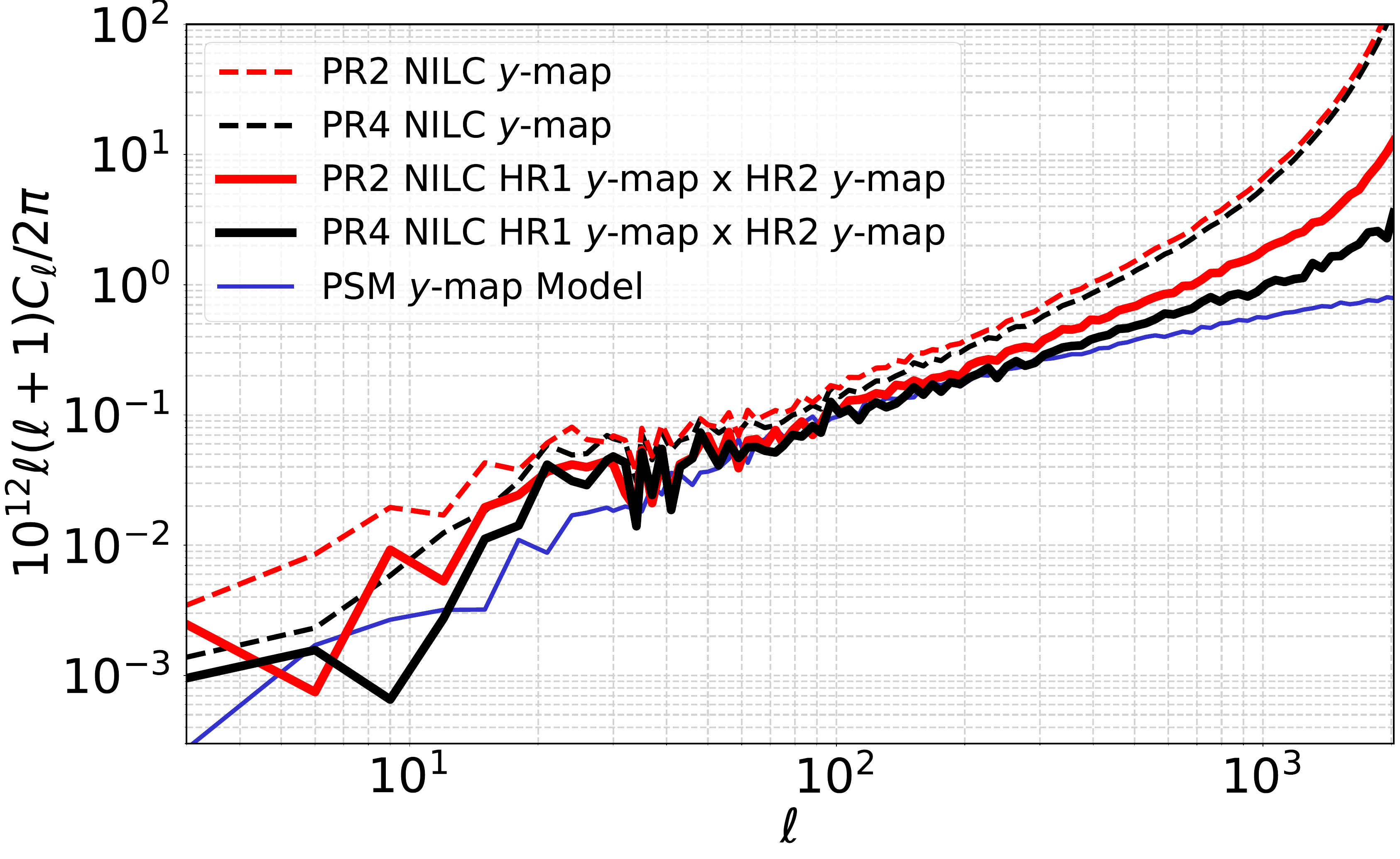}
\includegraphics[scale=0.11]{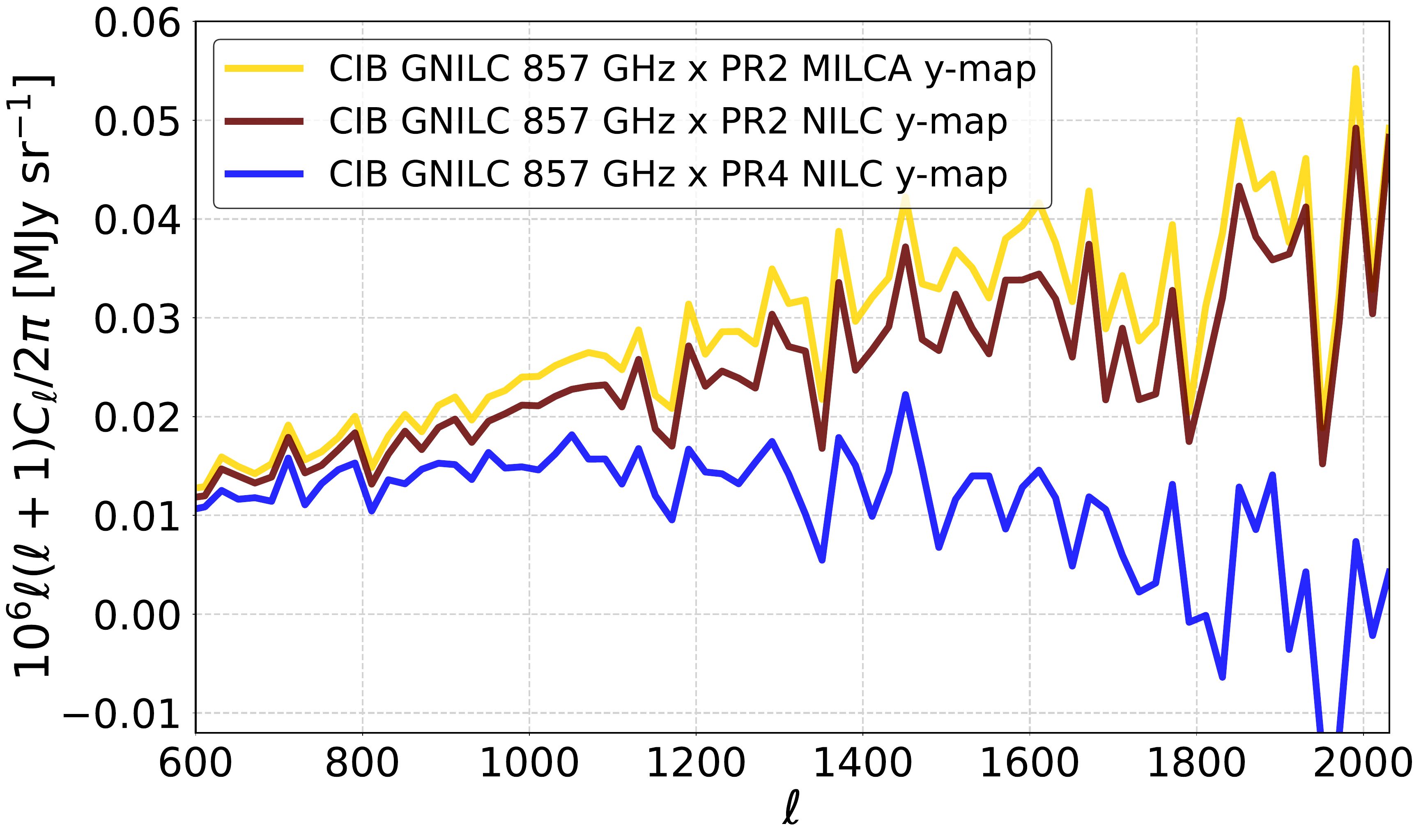}
\caption{\emph{Left}: The angular power spectra of the PR4 NILC $y$-map (black) and PR2 NILC $y$-map (red), including their auto-power spectra (dashed) and cross-power spectra of HR1 and HR2 $y$-maps (solid) for noise bias correction. \emph{Right}: The cross power spectra of the PR4 NILC (blue), PR2 NILC (red) and PR2 MILCA (yellow) $y$-maps with the \textit{Planck} GNILC $857$\,GHz CIB template.}
\label{fig:cls}       
\end{figure}

The angular power spectrum of the thermal SZ $y$-map is as a robust cosmological probe \cite{sz_power_spectrum}. Benefiting from the extensive sky coverage provided by the \textit{Planck} $y$-map, we can glean information across a broad range of angular scales, covering emissions from diffuse hot gas to compact galaxy clusters. Figure~\ref{fig:cls} shows the power spectra of the PR4/PR2 $y$-maps (\emph{left}, black/red dashed) over $56\%$ of the sky in the multipole range $\ell = 2$ to $2048$, after correcting for the mask, beam, and pixel windows. The PR4 $y$-map exhibits lower power than the PR2 $y$-map at small angular scales primarily due to reduced noise levels. 
To correct for the noise bias in the power spectra, we also compute the cross-power spectrum between the HR1 and HR2 $y$-maps (solid lines), as their noise is largely uncorrelated.


After correcting for the noise bias, the PR4 $y$-map (solid black) shows consistent reconstruction at intermediate multipoles ($\ell\sim$ 30-300) and lower power at high multipoles compared to the PR2 $y$-map (solid red). This difference at high multipoles is due to lower residual contamination from extragalactic foregrounds, in particular from CIB, as demonstrated hereafter. 

To assess residual contamination from CIB, cross-power spectra between the various $y$-maps and the \textit{Planck} GNILC CIB template at $857$\,GHz \cite{GNILC_CIB:Plk2015XLVIII} were computed over $50\%$ of the sky. The $857$\,GHz CIB template was chosen because this frequency channel was not used for reconstructing the $y$-maps at multipoles above $\ell= 300$, preventing spurious correlations between data products due to shared frequency channels. Results in the multipole range $\ell=600$ to $2000$ are shown in Figure~\ref{fig:cls} (\emph{right}). The PR4 NILC $y$-map (solid blue) exhibits significantly lower CIB contamination compared to the PR2 NILC (solid red) and PR2 MILCA (solid yellow) $y$-maps. On average, the PR4 $y$-map demonstrates a $34\%$ decrease in residual CIB contamination across the multipole range $\ell=600$-$2048$. As the noise variance is much higher than the CIB variance, even a slight reduction in noise in the PR4 data set enables NILC to effectively remove a substantial portion of CIB contamination at small angular scales, resulting in the observed improvements for the PR4 $y$-map.

\section{Conclusions} \label{sec:conclusion}

We have presented a new all-sky Compton $y$-parameter map using the \textit{Planck} PR4 data, which shows significant improvements over the \textit{Planck} PR2 $y$-maps. Through various checks in the pixel and harmonic domains at the map, 1-PDF and power spectra levels, we showed that the PR4 $y$-map has reduced contamination from instrumental noise ($\sim$ 7\%), CIB ($\sim$ 34\%), and Galactic dust compared to the PR2 $y$-maps. 

The \textit{Planck} PR4 NILC $y$-map, the half-ring (HR1 and HR2) $y$-maps, and the associated masks are available for public use at \url{https://doi.org/10.5281/zenodo.7940376} and in the Planck Legacy Archive (\url{http://pla.esac.esa.int/pla}).

\section*{Acknowledgements} \small{We thank Marcos López-Caniego for providing point-source masks for PR4 data. JC acknowledges financial support from the Concepción Arenal Programme of the Universidad de Cantabria. MR and JC acknowledge financial support from the CSIC programme ’Ayuda a la Incorporación de Científicos Titulares’ provided under the project 202250I159. Some of the results in this paper have been derived using the healpy \cite{healpy} and HEALPix \cite{HEALPix} packages. }

\end{document}